\documentclass[12pt]{article}
\usepackage{graphicx}
\usepackage{amsfonts}
\usepackage{amssymb,amsmath}
\usepackage{latexsym}
\usepackage{color}
\usepackage{cite}
\usepackage{bm}
\input{colordvi.tex}

\renewcommand{\thefootnote}{\#\arabic{footnote}}

\begin{document}

\renewcommand{\thepage}{\arabic{page}}
\setcounter{page}{1}
\renewcommand{\thefootnote}{\#\arabic{footnote}}

\begin{titlepage}

\begin{flushright}
ICRR-Report-675-2014-1\\
IPMU14-0088 
\end{flushright}

\begin{center}

\vskip .5in
{\Large \bf Isocurvature perturbations and tensor mode \vspace{2mm}  \\
in light of Planck and BICEP2 
}

\vskip .45in

{\large
Masahiro~Kawasaki$^{1,2}$,
Toyokazu~Sekiguchi$^3$,
Tomo~Takahashi$^4$ \vspace{2mm}  \\ 
and 
Shuichiro~Yokoyama$^1$
}

\vskip .45in

{\em
$^1$
Institute for Cosmic Ray Research, University of Tokyo, Kashiwa 277-8582, Japan \vspace{2mm} \\
$^2$
Kavli Institute for the Physics and Mathematics of the Universe (Kavli IPMU), \\
TODIAS, the University of Tokyo, 5-1-5 Kashiwanoha, Kashiwa, 277-8583, Japan  \vspace{2mm}  \\
$^3$
Helsinki Institute of Physics, University of Helsinki, PO Box 64, FIN-00014,
Finland  \vspace{2mm} \\
$^4$
Department of Physics, Saga University, Saga 840-8502, Japan
}

\end{center}

\vskip .4in

\begin{abstract}
We investigate the degeneracy of the isocurvature perturbations and the primordial gravitational waves, by using recent observations of the cosmic microwave background (CMB)
reported by Planck and BICEP2 collaborations. 
We show that the tension in the bound on the tensor-to-scalar ratio $r$ between Planck and BICEP2 
can be resolved by introducing the anti-correlated isocurvature perturbations.
Quantitatively, we find that with the anti-correlated isocurvature perturbations 
the constraints on $r$ from Planck alone and BICEP2 results can be consistent at 68 \% C.L.

\end{abstract}
\end{titlepage}

\setcounter{footnote}{0}

\section{Introduction}

Recently 
the detection of primordial gravitational waves has been reported by BICEP2~\cite{Ade:2014xna,Ade:2014gua}. 
The size of the gravitational waves is usually characterized by the so-called tensor-to-scalar ratio 
$r$ 
and its reported value from BICEP2 is
\begin{equation}
r = 0.2^{+0.07}_{-0.05} \qquad (68~\%~{\rm C.L.}),
\end{equation}
which excludes $r=0$ with 7$\sigma$~\cite{Ade:2014xna}. However, this large value of $r$ is in strong tension with 
the temperature data from Planck in the framework of the standard $\Lambda$CDM+$r$ model which gives the limit $r < 0.11$ \cite{Ade:2013uln}.
Some possible extensions of the model to resolve the tension have been discussed since the data release of BICEP2 
such as the running spectral index~\cite{Ade:2014xna,Ma:2014vua,Czerny:2014wua}, the suppression of the adiabatic perturbations on large scales~\cite{Contaldi:2014zua,Miranda:2014wga,Abazajian:2014tqa,Hazra:2014aea}, 
the presence of the extra relativistic particles~\cite{Giusarma:2014zza,Zhang:2014dxk},
blue-tilted  tensor spectrum~\cite{Gerbino:2014eqa,Ashoorioon:2014nta,Wu:2014qxa},
the existence of anti-correlated isocurvature mode~\cite{Kawasaki:2014lqa,Bastero-Gil:2014oga},  
and so on. 
Among these possibilities, we in this paper focus on the model with isocurvature perturbations.
Since the constraints on $r$ from the temperature data come from the enhancement due to the tensor mode contribution 
on large scales, if one can reduce the power on large scales, a large tensor mode contribution can be accommodated, 
which can be realized by assuming the anti-correlated isocurvature mode since it 
suppresses the TT power spectrum on large scales.  Since such anti-correlated isocurvature perturbations can naturally arise 
in multiple scalar field models such as
the curvaton~\cite{Enqvist:2001zp,Lyth:2001nq,Moroi:2001ct}\footnote{
Isocurvature perturbations in the curvaton model have been discussed in~\cite{Moroi:2002rd,Lyth:2003ip,Beltran:2008aa,Moroi:2008nn,Takahashi:2009cx}.
}, this issue deserves a further study. 
In this paper, we investigate how the tension can be reduced  in a more quantitative way by 
performing Markov Chain Monte Carlo (MCMC) analysis using Planck,  BICEP2 and some other  data. 
To discuss  the effects of a broad class of models with isocurvature perturbations, we study not only a case with
anti-correlated isocurvature mode, but also a generally correlated one, as well as varying the isocurvature spectral index.
As will be discussed, the existence of the anti-correlated isocurvature mode can allow large contribution from the tensor mode 
consistent with Planck temperature data, which may give a hint of the existence of the isocurvature mode.

It should also be noted that constraints on isocurvature perturbations in the light of BICEP2 data are worth investigating on their own.
Although the constraints on isocurvature perturbations have been studied in many works, 
to our knowledge, there are only a few studies on isocurvature perturbations in the existence of the tensor 
mode~\cite{Kawasaki:2007mb,Valiviita:2012ub,Savelainen:2013iwa}.
Now the BICEP2 data indicates that the tensor mode should not be neglected, 
the investigation of isocurvature perturbations in the presence of the tensor mode  would be an important subject.

This paper is organized as follows. In the next section, we briefly give a formalism to 
discuss the isocurvature perturbations and the effects 
of isocurvature perturbations and tensor mode in the CMB power spectra. 
Then in Section~\ref{sec:analysis}, we describe the method of the analysis. In Section \ref{sec:result}, 
we show  our results and discuss how the tension between Planck and BICEP2 data can be 
resolved by introducing isocurvature perturbations.
The final section is devoted to the conclusion of this paper.

\section{Effects of isocurvature perturbations and the tensor mode}  \label{sec:iso}

\subsection{Definitions}

First we summarize the formalism to discuss isocurvature perturbations and set our notations.
Isocurvature perturbations for a component $i$ is usually defined as 
\begin{equation}
S_i \equiv 3 (\zeta_i - \zeta_r), 
\end{equation}
where $\zeta_i$ is the curvature perturbation on the uniform energy density hypersurface of $i$ and 
the subscript $r$ indicates radiation. Although various isocurvature modes have been discussed in the 
literature (see Ref. \cite{Ade:2013uln} and references therein),  we only consider cold dark matter (CDM) 
mode in this paper\footnote{
Since  the effects of 
the baryon isocurvature mode on the CMB power spectrum are identical with the CDM one, 
our results can also be translated into the baryonic mode by rescaling its amplitude in proportion to their energy densities.
}. Thus in the following, we denote $S_{\rm CDM}$ simply as $S$ 
and also the adiabatic curvature perturbations during radiation dominated era $\zeta_r$
simply as $\zeta$.
Depending on the model, the curvature and isocurvature perturbations can be correlated or uncorrelated. 

The power spectrum of the initial perturbations $P_{XY}(k)$ 
is defined by $\langle X(\vec k) Y(\vec k')\rangle=P_{XY}(k)(2\pi)^3\delta^{(3)} (\vec k+\vec k')$
with $X$ and $Y$ being either $\zeta$ or $S$.
To discuss the size and the scale-dependence of the curvature (adiabatic) and isocurvature perturbations, 
we characterize the power spectra for the auto- and cross-correlation of $\zeta$ and $S$ as 
\begin{eqnarray}
\mathcal{P}_\zeta (k) =  \frac{k^3}{2\pi^2} P_{\zeta\zeta} (k) = A_\zeta (k_0) \left( \frac{k}{k_0} \right)^{n_s - 1}, \\
\mathcal{P}_S (k)  =  \frac{k^3}{2\pi^2} P_{SS} (k) = A_S (k_0) \left( \frac{k}{k_0} \right)^{n_{\rm iso} - 1}, \\
\mathcal{P}_{\zeta S} (k) = \frac{k^3}{2\pi^2}  P_{\zeta S} (k) = A_{\zeta S} (k_0) \left( \frac{k}{k_0} \right)^{n_c - 1}, 
\end{eqnarray}
where $A_\zeta, A_S$ and $A_{\zeta S}$ are respectively the amplitude of the power spectra, 
$n_s$, $n_{\rm iso}$ and $n_c$ are spectral indices for the auto-correlation power spectra of $\zeta, S$ 
and the cross-correlation one, respectively.  We define these quantities at the reference 
scale $k_0$ which we adopt $k_0 = 0.05~{\rm Mpc}^{-1}$ for the analysis in this paper.

Depending on the model, the degree of the correlation between the adiabatic and isocurvature perturbations varies. 
To quantify the cross-correlation, we define the correlation parameter as
\begin{equation}
\gamma_{\rm iso} 
\equiv \frac{P_{\zeta S} (k_0)}{\sqrt{ P_\zeta (k_0) P_S (k_0)}}.
\end{equation}
We denote the cases with   $\gamma_{\rm iso} >0$, $\gamma_{\rm iso} <0$ 
and $\gamma_{\rm iso} =0$ as 
positively correlated,  negatively (anti-) correlated 
and uncorrelated isocurvature perturbations, respectively.
Although  the spectral indices can be generally treated as independent 
parameters, in this paper, we assume the case with $n_s = n_{\rm iso} = n_c$, which is motivated by 
the curvaton model where the adiabatic and isocurvature perturbations originate from a single scalar field.
When we consider the case with uncorrelated isocurvature mode, we also investigate a possibility of $n_s \ne n_{\rm iso}$
which is the case for the axion isocurvature model\footnote{
While in many models $n_{\rm iso}$ is close to 1, 
in some models $n_{\rm iso}$ can be as large as $n_{\rm iso} \simeq 4$~\cite{Kasuya:2009up}.
}.

In the analysis, to parameterize the size of isocurvature perturbations, the fraction of their contribution 
is commonly used and it is defined as 
\begin{equation}
\alpha_{\rm iso} 
\equiv \frac{P_S (k_0)}{P_\zeta (k_0) + P_S (k_0) }.
\end{equation}
Since we consider cases with isocurvature perturbations, 
we define the tensor-to-scalar ratio including the isocurvature one in the scalar part as
\begin{equation}
r \equiv {P_T (k_0) \over P_{\zeta} (k_0) + P_S (k_0)},
\end{equation}
where $P_T$ denotes the power spectrum of the tensor mode.
Besides, we assume a scale-invariant tensor power spectrum, 
i.e. $P_T(k)k^3={\rm const}$, throughout this paper.

\subsection{Effects on CMB}

\begin{figure}[htbp]
  \begin{center}
    \resizebox{110mm}{!}{
     \includegraphics{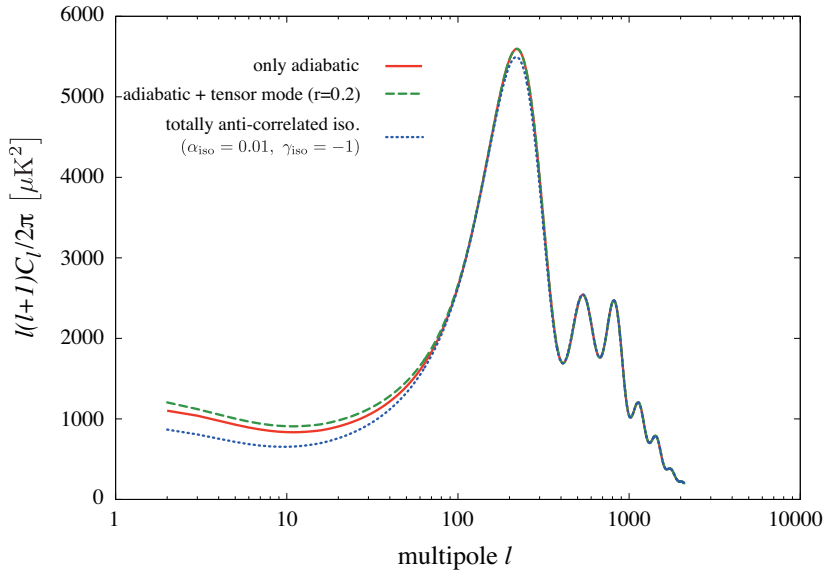}
}
  \end{center}
  \caption{CMB TT power spectrum for the cases with only adiabatic (red solid line), adiabatic and totally anti-correlated isocurvature ($\alpha_{\rm iso} = 0.01,~\gamma_{\rm iso} = -1$) (blue dotted line),
  and adiabatic and tensor modes with $r = 0.2$ (green dashed line).
  For other cosmological parameters, here we adopt the best-fit values obtained for a flat $\Lambda$CDM model 
  by Planck and WMAP 9-year polarization data~\cite{Ade:2013zuv}. 
  }
  \label{fig:cl}
\end{figure}

The contribution of the adiabatic and isocurvature modes 
to temperature fluctuations can be roughly 
evaluated by looking at those for the Sachs-Wolfe (SW) effect on large scales (see, e.g., Ref. \cite{Liddle:1993fq}):
\begin{equation}
\left. \frac{\Delta T}{T} \right|_{\rm SW} = - \frac15 \zeta - \frac25 S.
\end{equation}
Thus the CMB TT angular power spectrum from the SW effect can be written as 
\begin{equation}
\left. \frac{l (l+1) C^{\rm TT}_l }{2\pi}  \right|_{\rm SW} 
\propto 
\mathcal{P}_\zeta + 4 \mathcal{P}_{\zeta S} + 4 \mathcal{P}_S. 
\end{equation}
When the correlation parameter is $\gamma_{\rm iso}< 0$, $\mathcal{P}_{\zeta S}$ can give 
a negative contribution to the SW effect, which suppresses the power on large scales.
In Fig.~\ref{fig:cl}, CMB TT power spectra are shown for the cases with adiabatic, 
anti-correlated isocurvature 
and tensor modes. As shown in the figure, the spectrum for the anti-correlated isocurvature case is suppressed 
on large scale, which can cancel the tensor mode contribution. Thus we expect that the inclusion of 
the anti-correlated isocurvature mode would help to resolve the tension between the temperature data from Planck 
and the observation of B-mode polarization from BICEP2. 

In the next section, we study this issue in a more quantitative way by using a MCMC analysis
to fit the model with isocurvature perturbations and tensor mode. 

\section{Analysis}\label{sec:analysis}

In our analysis, we assume a flat $\Lambda$CDM model with isocurvature perturbations and a tensor mode, which consists of 
the  following cosmological parameters
\begin{equation}
(\Omega_bh^2,~\Omega_ch^2, H_0,~\tau,~n_s,~A_\zeta,~r,~\alpha_{\rm iso},~\gamma_{\rm iso},~n_{\rm iso},~n_c), 
\end{equation}
where $\Omega_bh^2$ and $\Omega_ch^2$ respectively are the
density parameters of baryon and CDM, $H_0=100h$ is the Hubble parameter in units of km/sec/Mpc, 
$\tau$ is the optical depth of reionization. 
However, we do not vary all of these parameters completely independently.
Instead, we often treat some of the parameters for the initial power spectrum
as fixed ones or related in some way (e.g., we assume $n_s = n_{\rm iso} = n_c$ for some analysis).

We adopt the latest version of {\tt CosmoMC}~\cite{Lewis:2002ah,Lewis:2013hha}, 
which allows us to sample the posterior distribution of the 
cosmological parameters as well as various nuisance parameters which account for the
contributions of foregrounds to CMB.

As CMB data, we adopt the temperature anisotropy from Planck 
(hereafter P13)~\cite{Ade:2013hta,Ade:2013kta,Ade:2013zuv}, 
the E-mode polarization one from WMAP 9-year 
(WP9)~\cite{Bennett:2012zja,Hinshaw:2012aka} and
the E- and B-mode polarization ones from BICEP2 (B2)~\cite{Ade:2014xna,Ade:2014gua}\footnote{\
Note that we combine the polarization data of WP9 and B2 by simply 
adding $\chi^2$ from the likelihood codes of them. This is 
justified since these two data are dominated by the measurement noise
and the sample variance is subdominant.
}.
In some cases we also combine the measurements 
of the baryon acoustic oscillation scales \cite{Beutler:2011hx,Anderson:2012sa,Padmanabhan:2012hf} (BAO)
and the direct measurement of the Hubble parameter \cite{Riess:2011yx} (H0).

\section{Result}\label{sec:result}

Now we present our result. In Table \ref{tab:constraints}, 
we summarize constraints on cosmological parameters
related to the tensor and isocurvature perturbations for various
models, which we are to describe in the following.

\begin{table}
\begin{center}
\begin{tabular}{ll|rrr}
\hline \hline 
model & data & \multicolumn{3}{c}{parameter} \\
\hline \hline 
purely adiabatic & & $r$ & & \\
& P13+WP9 & $[0,~0.12]$ & & \\
& \quad+B2 & $[0.096,~0.24]$ & & \\
& \quad+B2+BAO+H0 & $[0.099,~0.24]$ & & \\
\hline
uncorrelated & & $r$ & $\alpha_{\rm iso}$ & \\
& P13+WP9 &$ [0,~0.11]$ & $[0,~0.030]$ & \\
& \quad+B2 & $[0.092,~0.23]$  & $[0,~0.024]$ & \\
& \quad+B2+BAO+H0 & $[0.095,~0.23]$  & $[0,~0.023]$ & \\
\hline
totally correlated & & $r$ & $\alpha_{\rm iso}$ & \\
& P13+WP9 & $[0,~0.10]$ & $[0,~0.0026]$ & \\
& \quad+B2 & $[0.098,~0.24]$ & $[0,~0.0015]$ & \\
& \quad+B2+BAO+H0 & $[0.10,~0.24]$ & $[0,~0.0015]$ & \\
\hline
totally anti-correlated & & $r$ & $\alpha_{\rm iso}$ & \\
& P13+WP9 & $[0,~0.17]$ & $[0,~0.012]$ & \\
& \quad+B2  & $[0.11,~0.26]$ & $[0,~0.015]$ & \\
& \quad+B2+BAO+H0  & $[0.12,~0.27]$ & $[0,~0.014]$ & \\
\hline
generally correlated & & $r$ & $\alpha_{\rm iso}$ & $\gamma_{\rm iso}$ \\
& P13+WP9+B2 & $[0.10,~0.25]$ & $[0,~0.028]$ & $[-1,~-0.38]$ \\
& \quad+BAO+H0 & $[0.11,~0.26]$ & $[0,~0.028]$ & $[-1,~-0.39]$ \\
\hline
varying $n_{\rm iso}$ (uncorrelated) & & $r$ & $\alpha_{\rm iso}$ & $n_{\rm iso}$ \\
& P13+WP9+B2 & $[0.073,~0.21]$ & $[0,~0.40]$ & $ [1.9,~4]$ \\
& \quad+BAO+H0 & $[0.072,~0.21]$ & $[0,~0.35]$ & $ [1.9,~4]$ \\
\hline\hline
\end{tabular} 
\caption{Constraints on cosmological parameters related to the tensor and isocurvature perturbations.
Shown are the 95 \% intervals of 1D marginalized posterior distributions for various model and
combinations of data sets. 
}
\label{tab:constraints}
\end{center}
\end{table}

\begin{figure}[htbp]
  \begin{center}
    \resizebox{80mm}{!}{\includegraphics{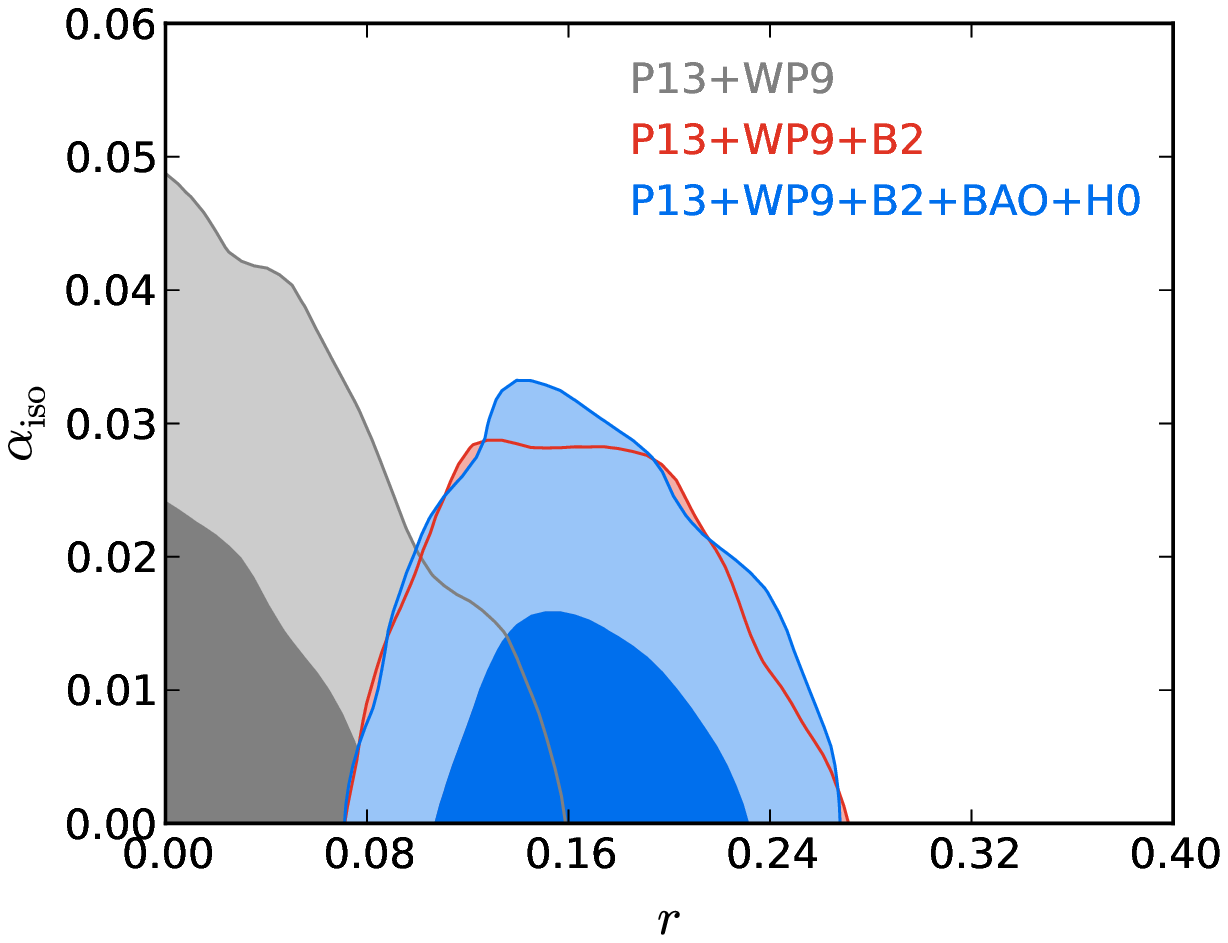}} \\
    \resizebox{80mm}{!}{\includegraphics{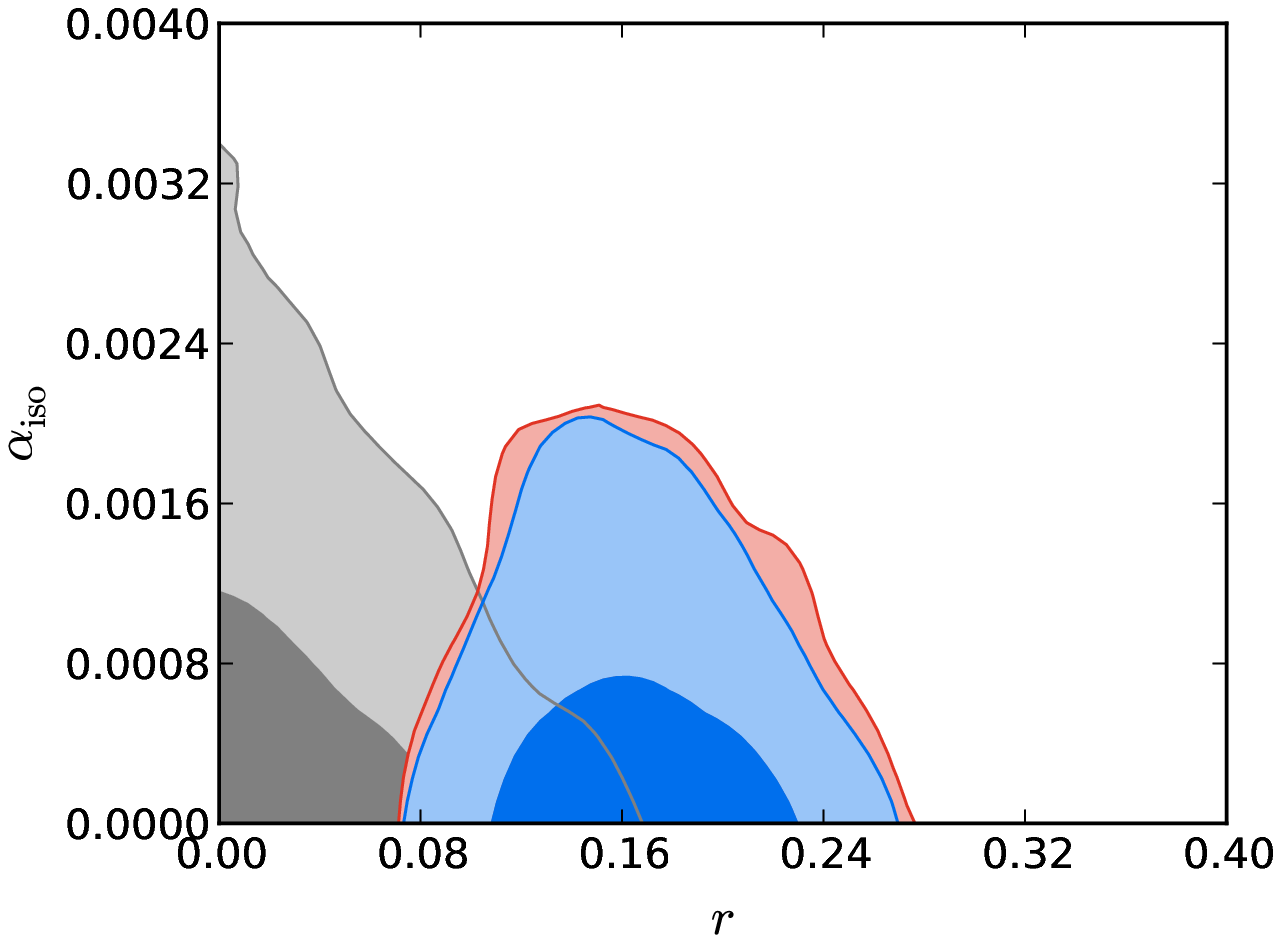}} \\
    \resizebox{80mm}{!}{\includegraphics{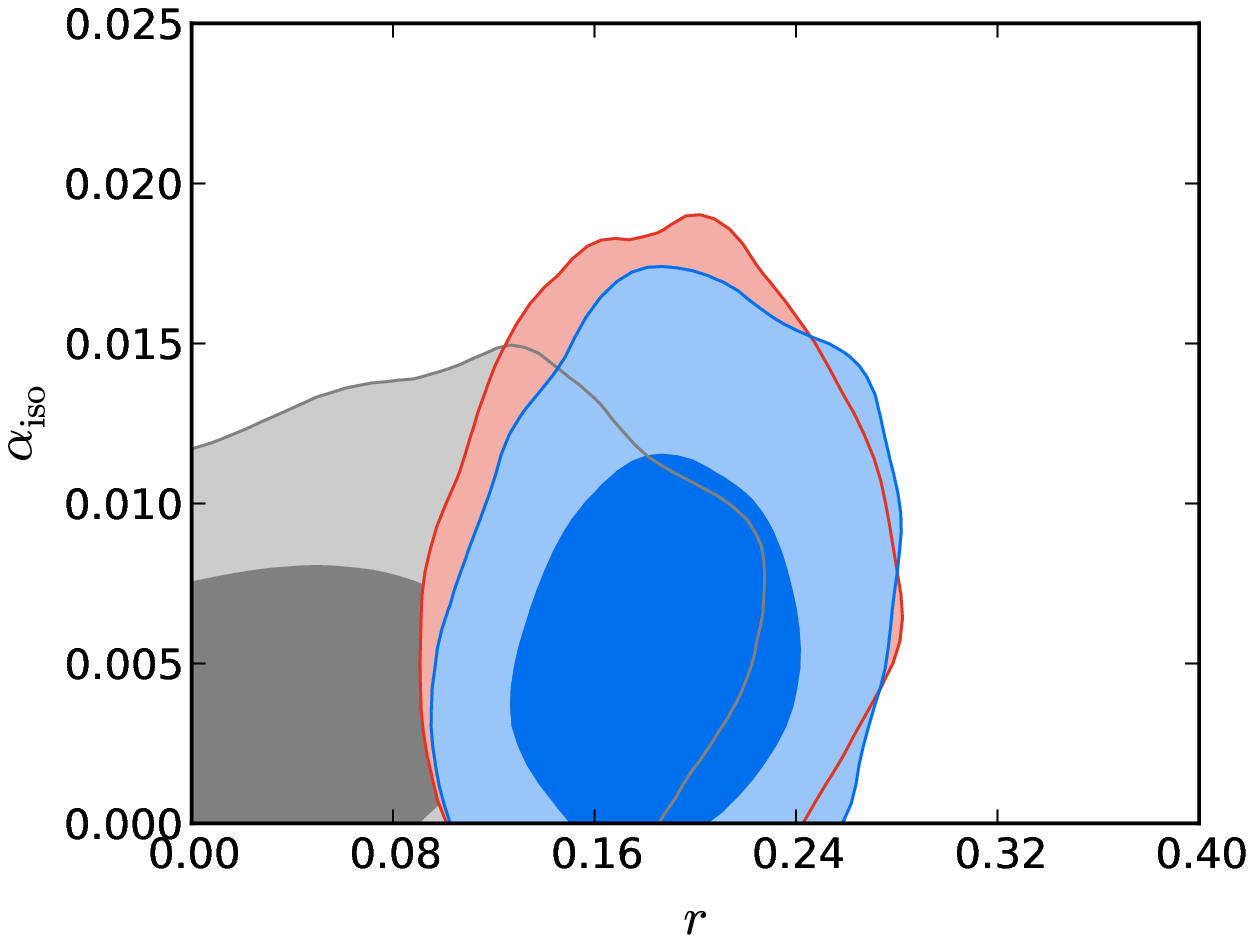}}
  \end{center}
  \caption{2D Constraints (68\% and 95\% C.L.) in the $r$-$\alpha_{\rm iso}$ plane for the cases of 
  fixed $\gamma_{\rm iso}=0$ (top), $1$ (middle) and $-1$ (bottom).
  Note that scales in $y$-axis differ significantly.
  }
  \label{fig:fixed_gamma}
\end{figure}

\begin{figure}[htbp]
  \begin{center}
    \resizebox{150mm}{!}{\includegraphics{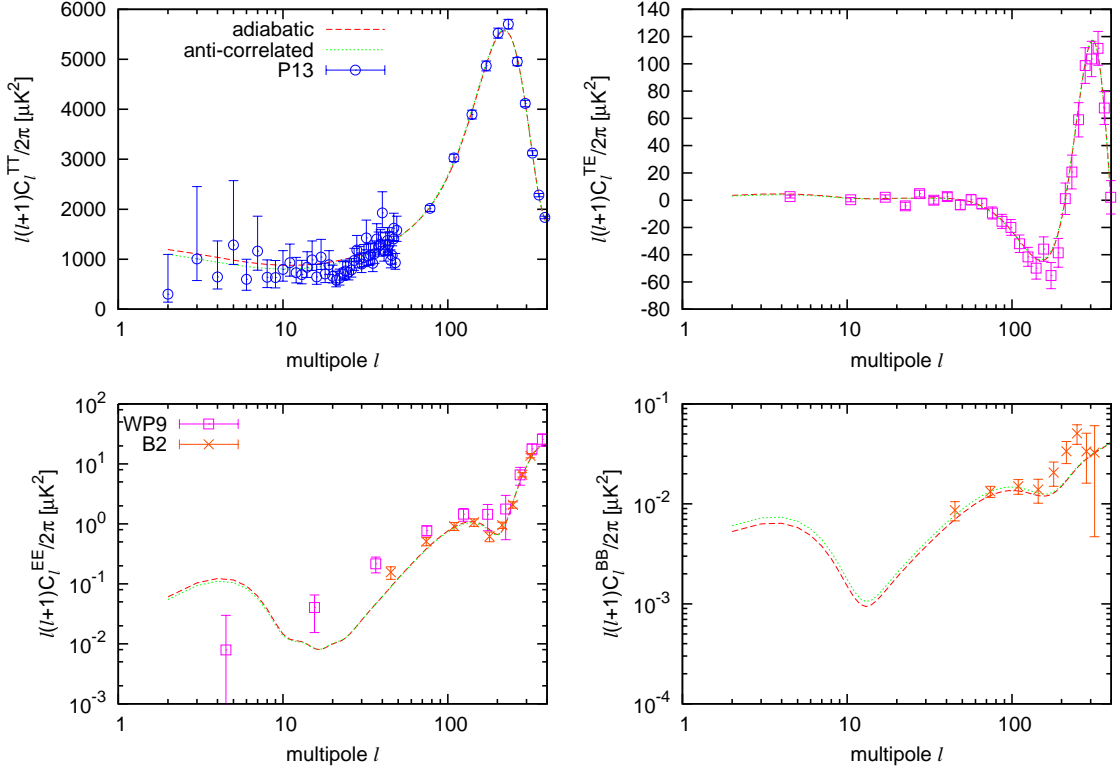}} 
  \end{center}
  \caption{Best-fit lensed CMB power spectra of the adiabatic (dashed red) and anti-correlated isocurvature (dotted green) models 
  to the data P13+WP9+B2+BAO+H0. The TT (top left), TE (top right), EE (bottom left) and BB (bottom right) power spectra are shown.
  Data points depicted are P13 (TT), WP9 (TE, EE) and B2 (EE, BB).
  Note that the $y$-axises in the top and bottom panels are shown in linear and logarithmic scales, respectively.
  }
  \label{fig:best_cl}
\end{figure}

Let us first see the constraints on isocurvature models with 
definite correlations $\gamma_{\rm iso}=0$, $1$ and $-1$, which
we refer to as uncorrelated, totally correlated and totally anti-correlated cases, respectively. 
Here we assumed $n_{\rm iso} = n_c = n_s $. 
In Fig.~\ref{fig:fixed_gamma}, we show constraints in the $r$-$\alpha_{\rm iso}$ plane
obtained from three combinations of data sets P13+WP9, P13+WP9+B2, and P13+WP9+B2+BAO+H0.
This figure enables us to see what types of the correlation of the isocurvature perturbations parameterized by $\gamma_{\rm iso}$
can weaken the discrepancy in the constraints on $r$ between the Planck and BICEP2 data.
In the totally anti-correlated case, there arise an overlap between
the allowed regions of P13+WP9 and P13+WP9+B2 at 68 \%~C.L.,  and a large overlap at  95\%~C.L. level.
Besides, we also computed the minimum values of $\chi^2$ in both the purely adiabatic and anti-correlated isocurvature models.
We found that inclusion of anti-correlated isocurvature perturbations
can significantly improve the fit to data; the minimum $\chi^2$ decreases 
by 6.6 (5.8) for the data set P13+WP2+B2 (P13+WP2+B2+BAO+H0).
These results indicate that,  in the presence of totally anti-correlated isocurvature perturbations, 
the tension between the Planck and BICEP2 data can be reduced.
In Fig.~\ref{fig:best_cl}, we compare the best-fit CMB power spectra of the purely adiabatic
and anti-correlated cases to the data P13+WMAP9+B2+BAO+H0.
We note that in these two best-fit models $(\alpha_{\rm iso},~r)$ takes the values (0,~0.17) and (0.0036,~0.18).
We can see the anti-correlated isocurvature model can give 
a lower TT power at large angular scales and a better fit to the Planck data than the purely adiabatic model.

\begin{figure}[htbp]
  \begin{center}
    \resizebox{140mm}{!}{\includegraphics{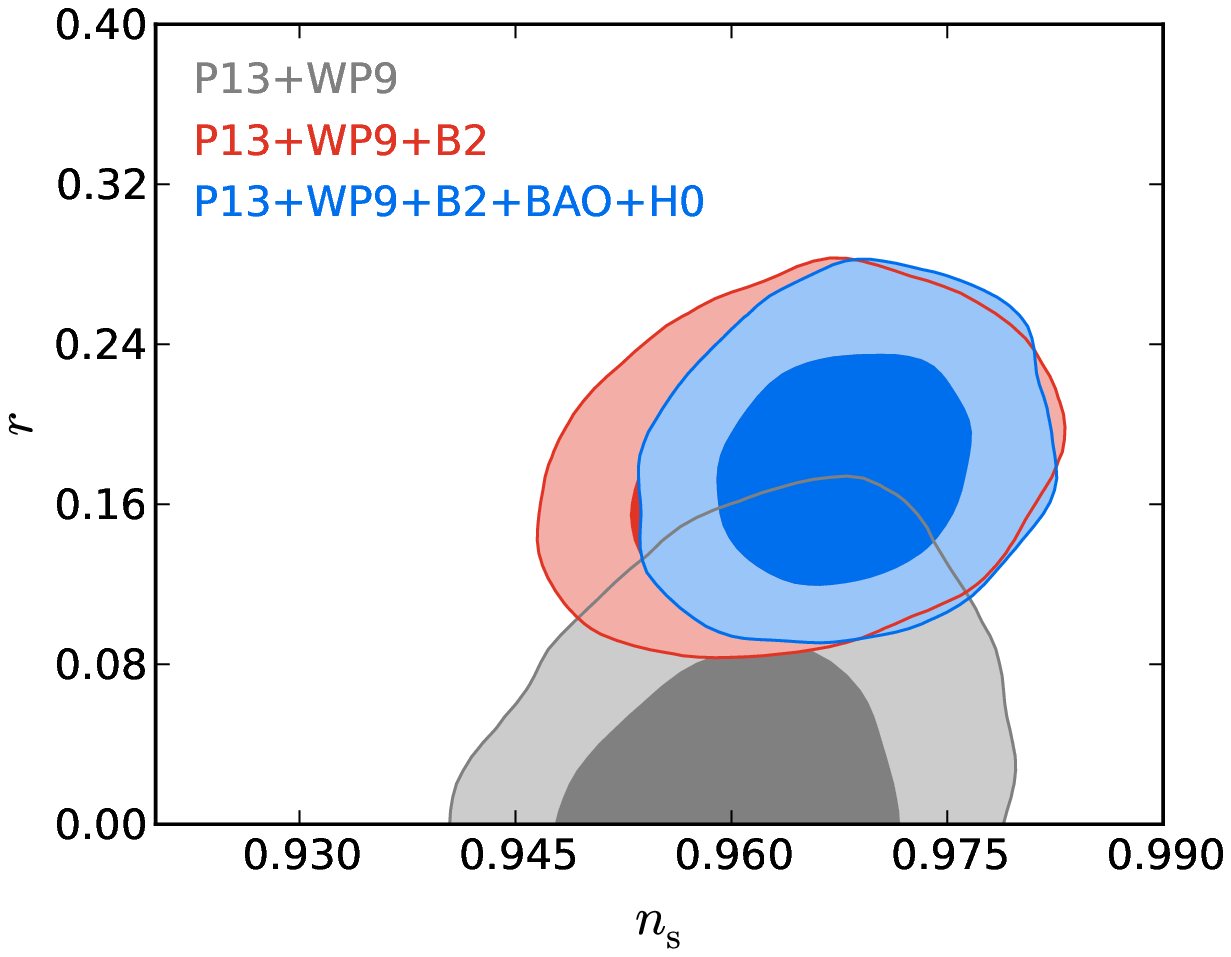}\includegraphics{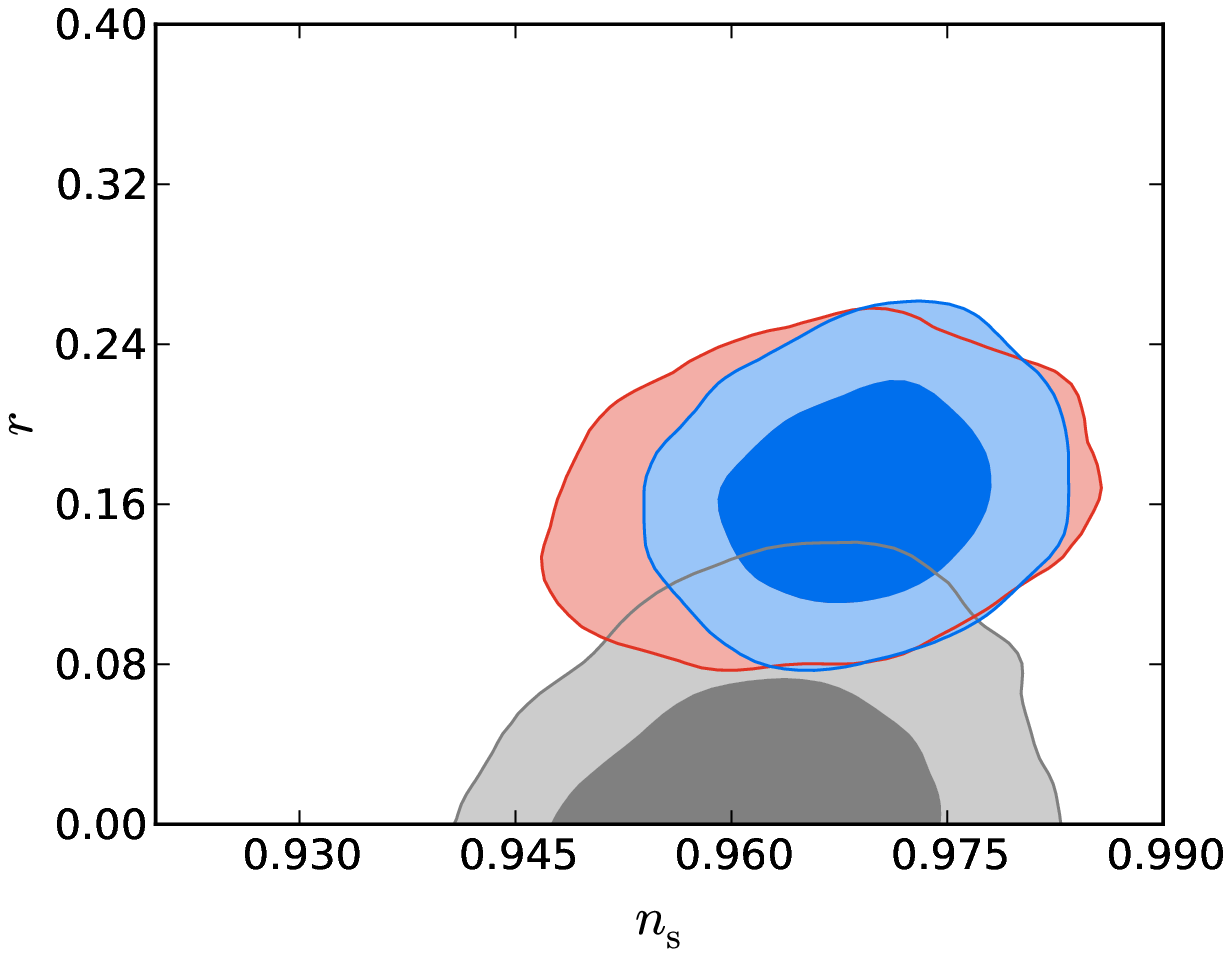}} \\
    \resizebox{140mm}{!}{\includegraphics{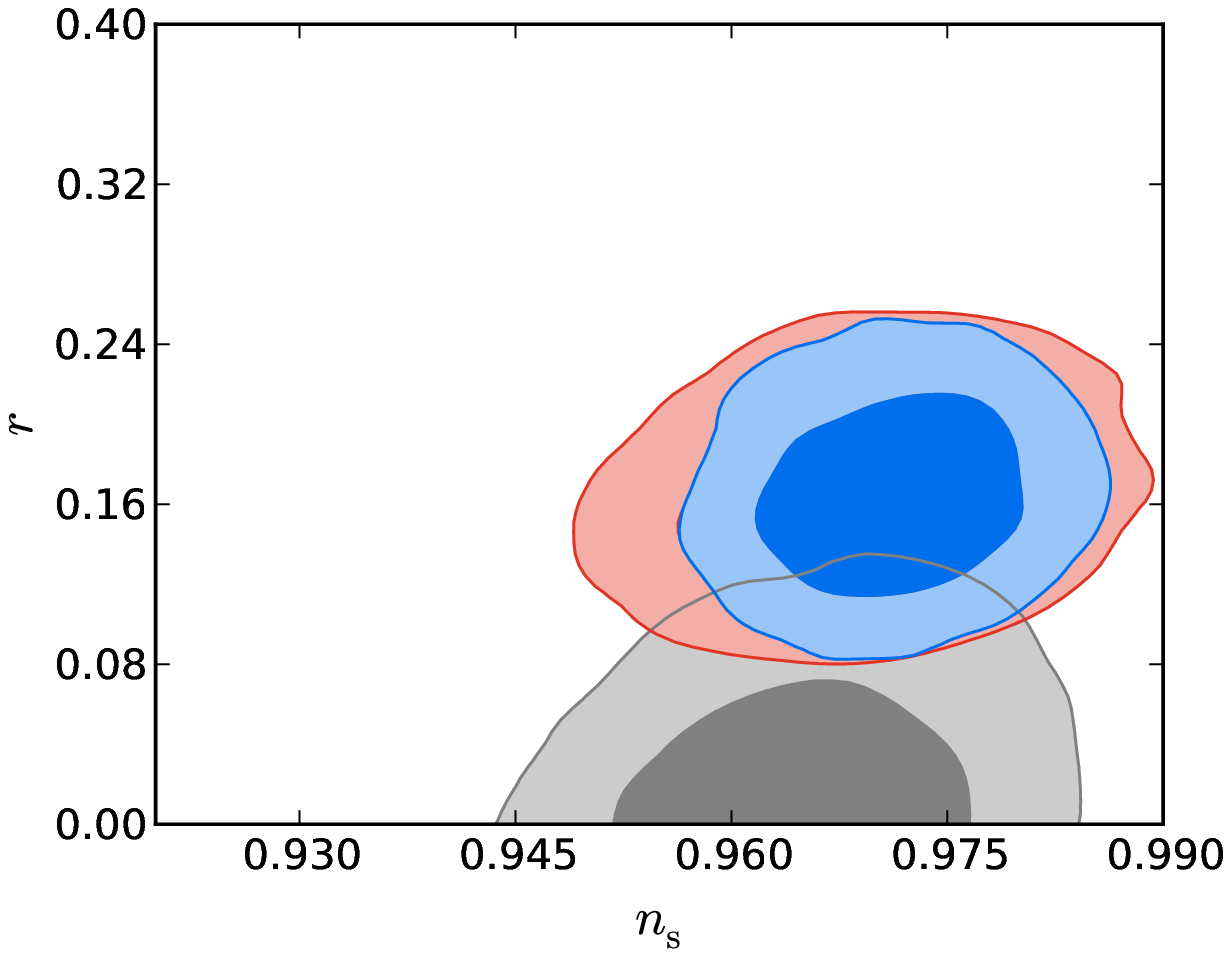}\includegraphics{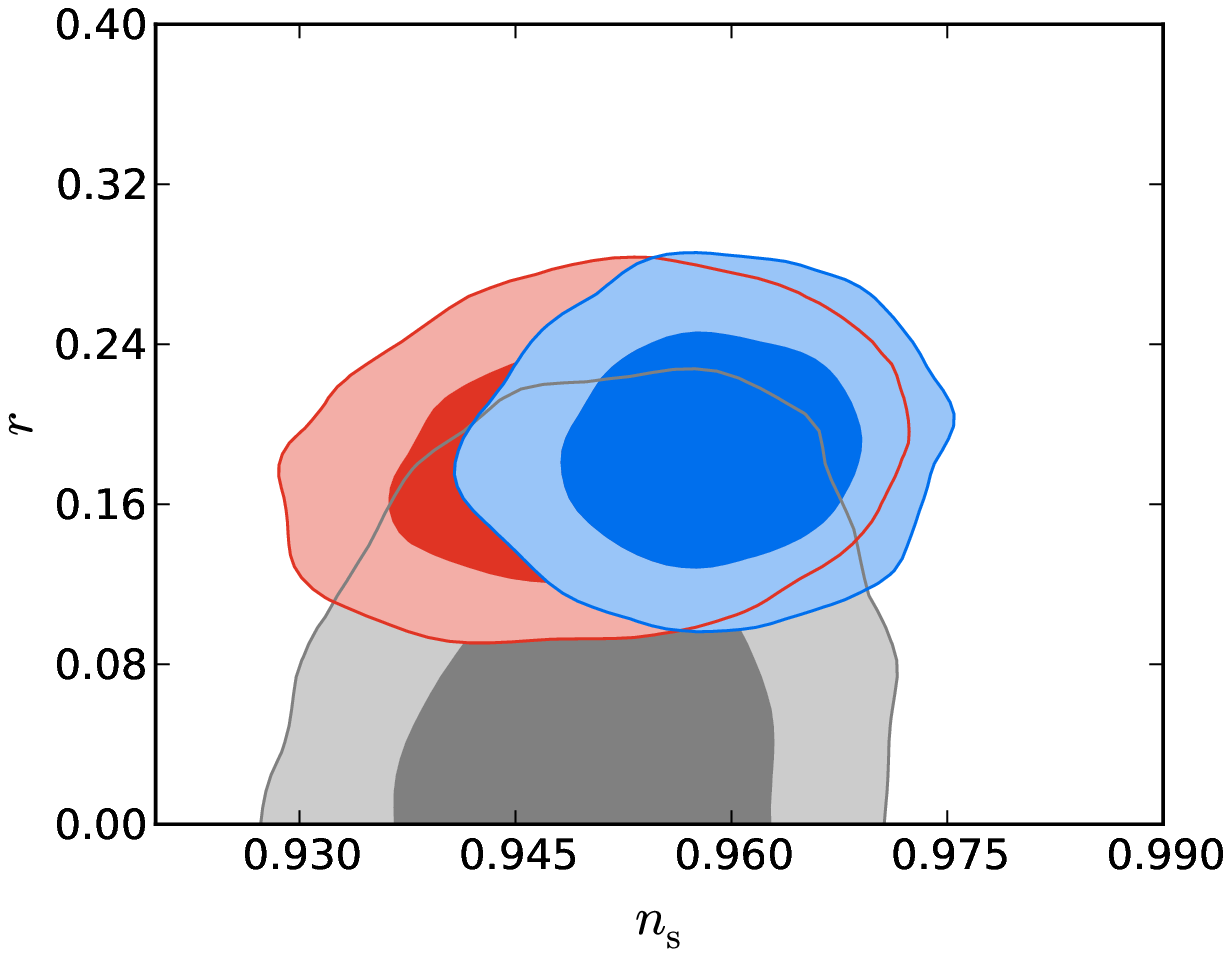}} \\
  \end{center}
  \caption{2D Constraints in the $n_s$--$r$ plane for the cases without
  isocurvature perturbations (top left), with isocurvature ones for 
  $\gamma_{\rm iso}=0$ (top right), $1$ (bottom left) and $-1$ (bottom right).
  }
  \label{fig:ns_r}
\end{figure}

On the other hand, in the case of uncorrelated or totally (positively) correlated isocurvature mode, 
allowed regions at 68 \% C.L. for data with and without BICEP2 do not overlap with each other
(see Fig.~\ref{fig:fixed_gamma}).
We also found that in these cases, the best-fit points to data are obtained near $\alpha_{\rm iso}=0$ and 
there are no significant reductions of the minimum $\chi^2$ from the purely adiabatic case
for all the data sets we adopted.

In Fig.~\ref{fig:ns_r}, we also show the constraints on the $n_s$--$r$ plane 
for the cases with no isocurvature (only adiabatic) perturbations, totally (positively) correlated, anti-correlated and uncorrelated 
isocurvature ones. 
The existence of anti-correlated isocurvature perturbations makes the Planck data to be consistent with larger tensor mode indicated by BICEP2.
This figure also clearly shows that the tension between Planck and BICEP2 can be weakened by 
introducing the anti-correlated isocurvature fluctuations.

\begin{figure}[htbp]
  \begin{center}
    \begin{tabular}{cc}
    \resizebox{70mm}{!}{\includegraphics{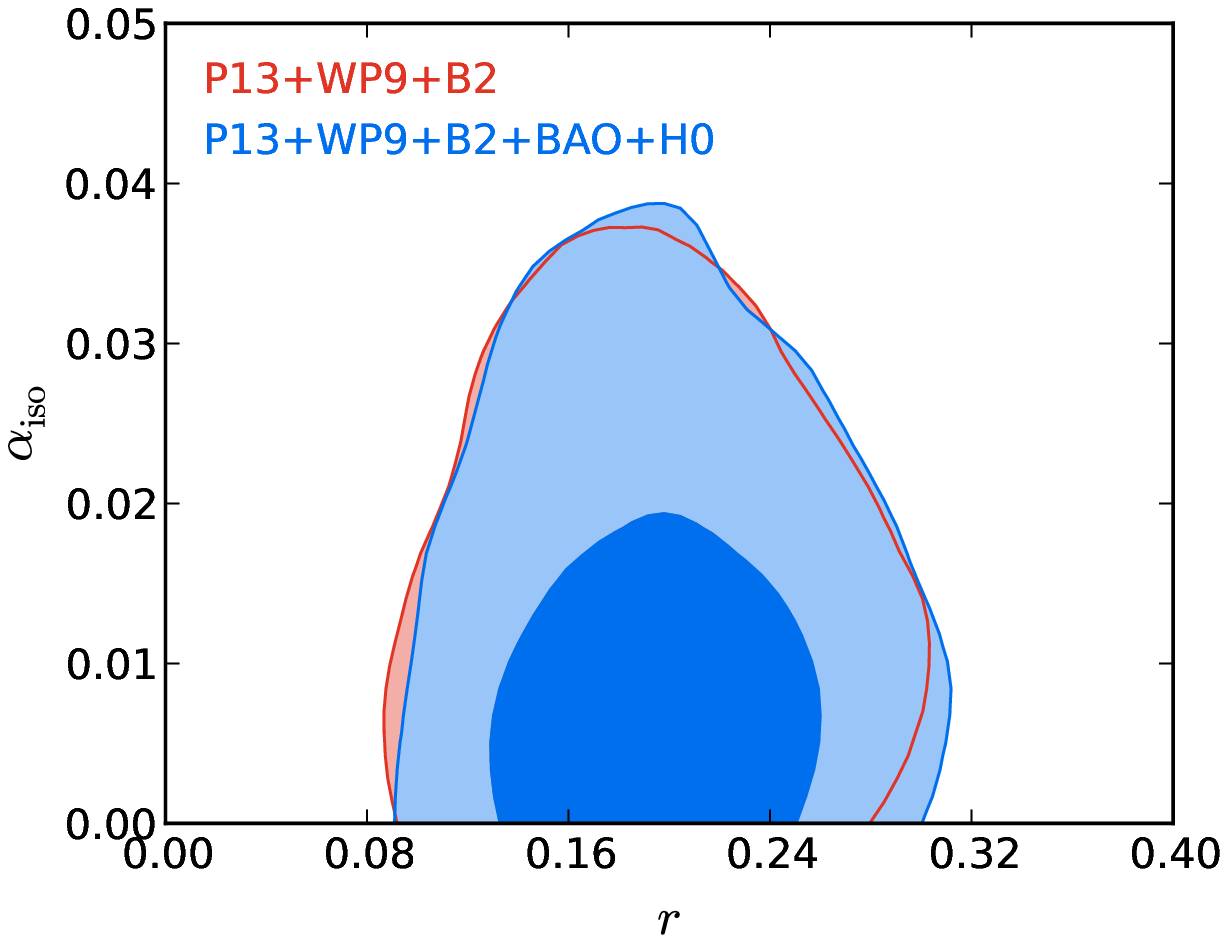}} &
    \resizebox{70mm}{!}{\includegraphics{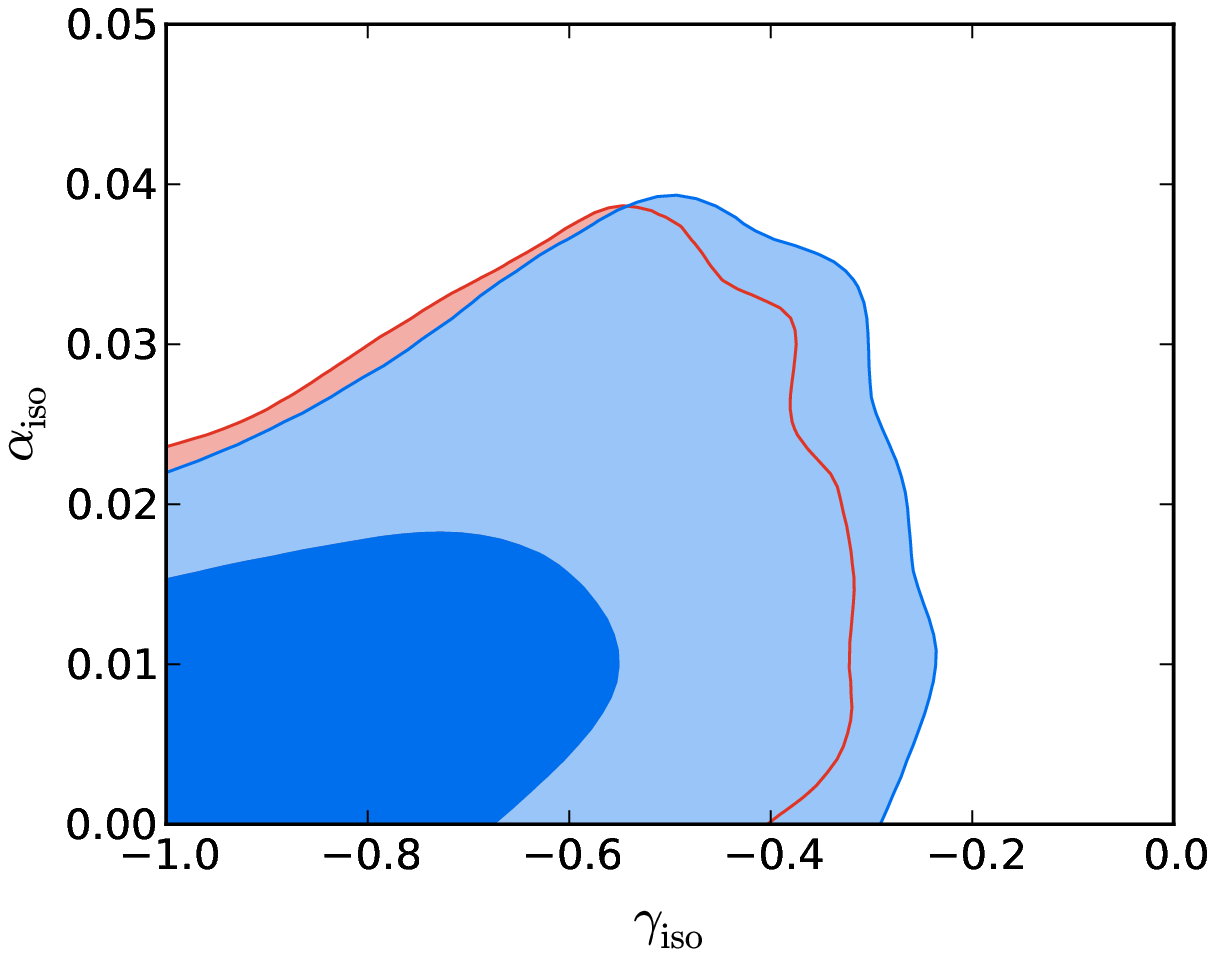}} 
  \end{tabular}
  \end{center}
  \caption{2D constraints in $r$-$\alpha_{\rm iso}$ and $\gamma_{\rm iso}$-$\alpha_{\rm iso}$ planes
  for the generally correlated case. 
  }
  \label{fig:varied_gamma}
\end{figure}

Next, we consider a case of generally correlated isocurvature perturbations 
(hereafter we denote as ``generally correlated case"), where $\gamma_{\rm iso}$ can vary in the range  $[-1,1]$.
Here we again assume $n_{\rm iso} = n_c =n_s$. 
In Fig.~\ref{fig:varied_gamma}, we show constraints in the $r$-$\alpha_{\rm iso}$ 
and the $\gamma_{\rm iso}$-$\alpha_{\rm iso}$ planes. 
From the figure, one clearly sees there is a strong preference for $\gamma_{\rm iso}=-1$, 
which is exactly what we expect from the previous result. 
In addition, we found that the best-fit points of the generally correlated case
are located near $\gamma_{\rm iso}=-1$ and there are no significant changes
in the minimum value of $\chi^2$ from the totally anti-correlated case.
However, this does not mean $\gamma_{\rm iso}$ can not differ significantly from $-1$ and
actually $\gamma_{\rm iso}\simeq-0.4$ is still allowed, though in such a case the improvement in the fit 
from the purely adiabatic case should be somewhat less prominent.
This is of particular importance when one considers a model with 
anti-correlated isocurvature and tensor perturbations.
In fact, to generate isocurvature perturbations, the existence of a light scalar field other than the inflaton is required.
In such a case,  isocurvature perturbations of the light scalar field can not correlate with the inflaton ones.
Thus, to obtain largely correlated isocurvature perturbations, the contribution from the other scalar field to the adiabatic perturbations
should be large.
Furthermore, when the correlation between adiabatic and isocurvature ones is relatively large, 
the tensor-to-scalar ratio tends to be suppressed than in the case where the adiabatic  perturbations
are sourced only from the inflaton, 
$r = r_{\rm inf}(1-\alpha_{\rm iso})(1-\gamma_{\rm iso}^2)$ ($r_{\rm inf}$: the tensor-to-scalar 
ratio for inflaton only).
Hence, in order to realize $| \gamma_{\rm iso} | \simeq 1$, we need to consider the situation where the inflaton contribution to the adiabatic curvature perturbations is relatively small and the tensor-to-scalar ratio is reduced.
We are able to construct a theoretical model based on e.g. the curvaton
scenario, which allows to generate enough anti-correlated isocurvature and tensor perturbations as discussed in Ref.~\cite{Kawasaki:2014lqa}.

So far we have seen that the presence of anti-correlated isocurvature perturbations 
can greatly reduce the tension between the Planck and BICEP2 data.
Then one may wonder if this means that nonzero isocurvature perturbations
are suggested by the combination of Planck and BICEP2.
However, this conclusion can not be drawn from the current data.
As is given in Table~\ref{tab:constraints} and also seen from Figs.~\ref{fig:fixed_gamma} 
and \ref{fig:varied_gamma}, the current data is still consistent with $\alpha_{\rm iso}=0$ 
at 95\% C.L.

\begin{figure}[htbp]
  \begin{center}
    \resizebox{70mm}{!}{\includegraphics{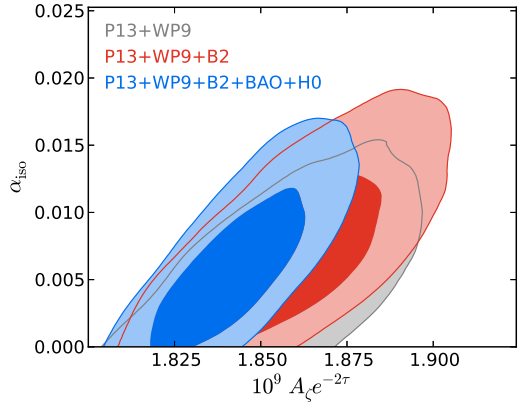}} 
  \end{center}
  \caption{Degeneracy between $A_\zeta e^{-2\tau}$ and $\alpha_{\rm iso}$ 
  in the anti-correlated isocurvature model.}
  \label{fig:clamp_alpha}
\end{figure}

Then how can we confirm the anti-correlated isocurvature model?
To discuss this issue, we look for combinations of cosmological parameters which significantly degenerate with $\alpha_{\rm iso}$ in the anti-correlated cases.
We found that $A_\zeta e^{-2\tau}$, which determines the amplitudes of the CMB power spectra (except for the tensor contribution), 
is one of the most degenerate quantities to $\alpha_{\rm iso}$, as shown in Fig.~\ref{fig:clamp_alpha}.
We expect precise measurements of the TE and EE power spectra can improve the determination
of $A_\zeta e^{-2\tau}$ and may pin down nonzero $\alpha_{\rm iso}$. 
In the near future, this will be achieved by the Planck polarization measurement \cite{Planck:2006aa}, 
which will also determine $\tau$ alone better and hence may help to break the parameter degeneracy further.
In addition, since the anti-correlated isocurvature model prefers larger $r$ compared to the purely adiabatic
model, precise measurements of the BB power spectrum would also strengthen (or weaken) the need
for anti-correlated isocurvature perturbations. Planck can measure $r$ with an accuracy $\Delta r\simeq$ 
a few percents~\cite{Planck:2006aa}. The precision  can be improved further by proposed
CMB surveys such as Lite-Bird~\cite{Matsumura:2013aja}, 
PIXIE~\cite{Kogut:2011xw} and PRISM~\cite{Andre:2013afa}, 
which will focus on the B-mode polarization at large angular scales.

\begin{figure}[htbp]
  \begin{center}
    \begin{tabular}{cc}
    \resizebox{70mm}{!}{\includegraphics{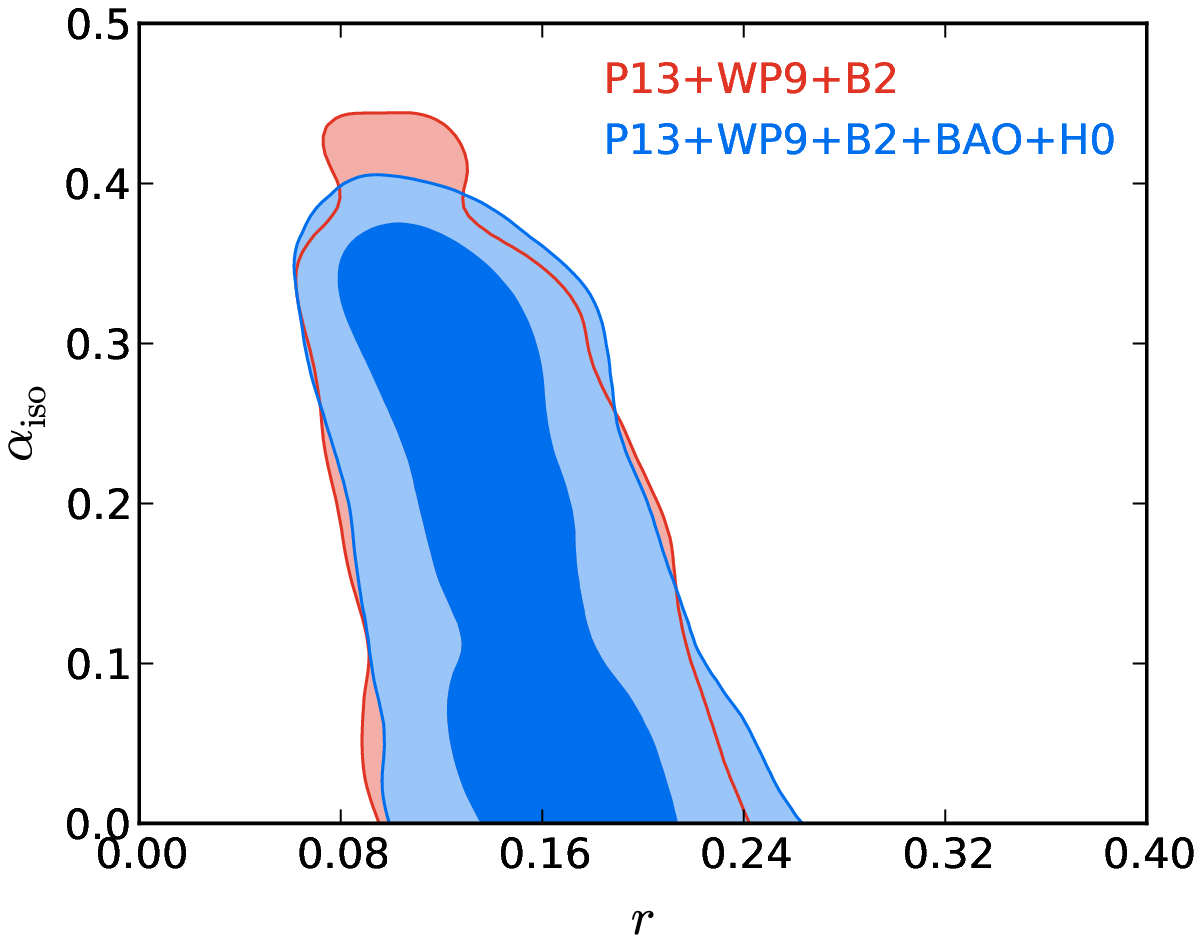}} &
    \resizebox{70mm}{!}{\includegraphics{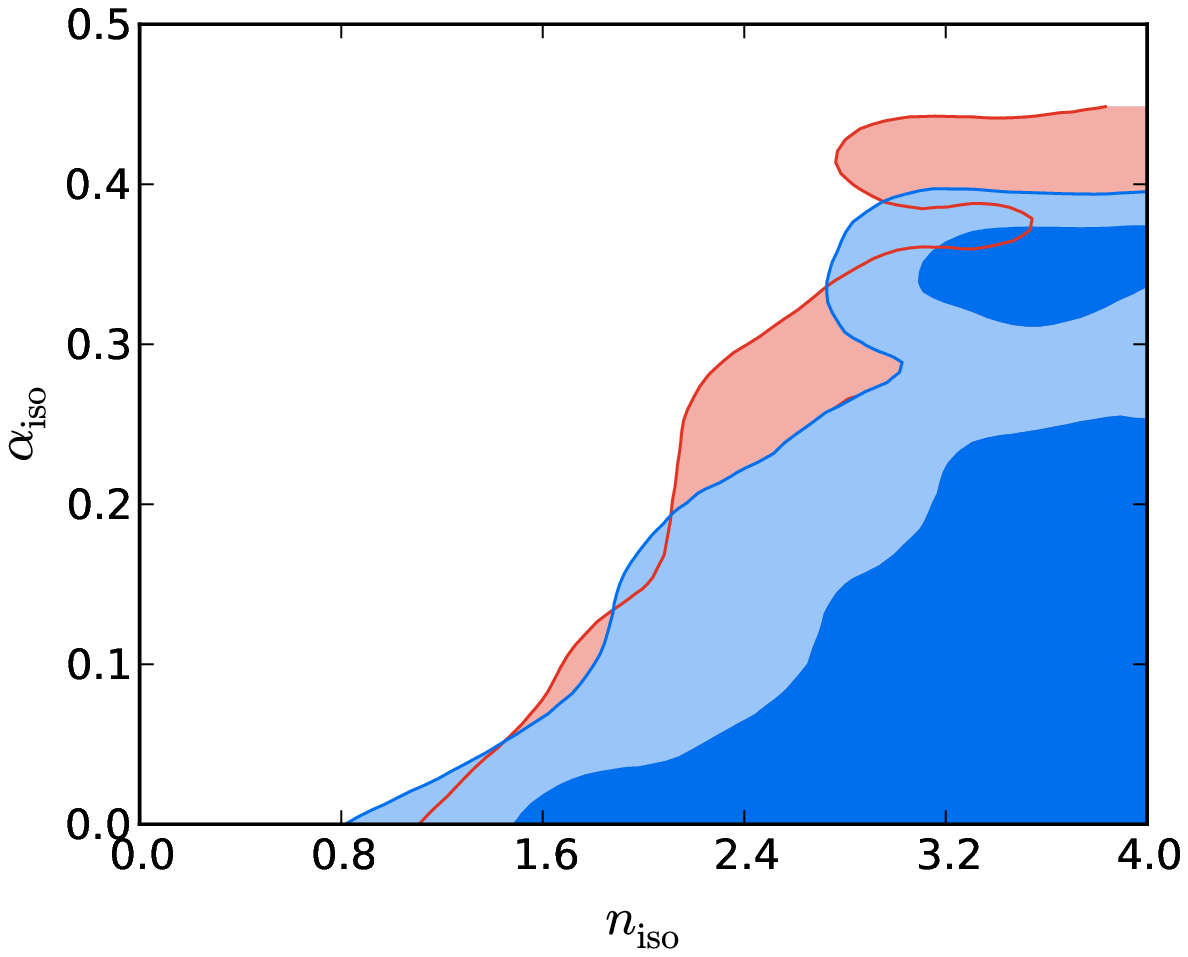}} 
  \end{tabular}
  \end{center}
  \caption{2D constraints in the $r$-$\alpha_{\rm iso}$ and the $n_{\rm iso}$-$\alpha_{\rm iso}$ planes
  for the uncorrelated isocurvature case with varying $n_{\rm iso}$. 
  }
  \label{fig:varied_niso}
\end{figure}

Finally we discuss a case of uncorrelated isocurvature perturbations $(\gamma_{\rm iso}=0)$ with $n_{\rm iso} \ne n_s$, 
which we refer to as  the varying $n_{\rm iso}$ case. 
Here $n_{\rm iso}$ can vary freely in the range $[0,4]$. 
Fig.~\ref{fig:varied_niso} shows the 2D constraints in the $r$-$\alpha_{\rm iso}$ and the $n_{\rm iso}$-$\alpha_{\rm iso}$ planes.
While a large fraction of isocurvature perturbations $\alpha_{\rm iso}\simeq0.4$ (at $k=0.05$~Mpc$^{-1}$)\footnote{
We note that such large blue isocurvature perturbations can affect the structure formation at small scales
and can be strongly constrained from reionization in spite of  large theoretical uncertainties \cite{Sekiguchi:2013lma}.
}
are allowed when the isocurvature spectrum is blue-tilted, we found no particular preference for
$\alpha_{\rm iso}\ne0$. 
On the other hand, we found no significant improvements in the fits to data from the purely adiabatic case.

\section{Conclusion}  \label{sec:conclusion}

In the conventional cosmology with adiabatic initial perturbations, 
there has been found a tension between the temperature data
from Planck and the B-mode one from BICEP2.
In this study, we have examined whether this tension can be mitigated
by introducing isocurvature perturbations.
We have found that, assuming the anti-correlated isocurvature perturbations, which
can be realized in e.g. the curvaton scenario, $\chi^2$ is
reduced by $\sim 6$ from the case of purely adiabatic perturbations
and the constraints on $r$ from Planck alone and
Planck combined with BICEP2 can be consistent at 68 \% C.L.
However, while the level of discrepancy can be reduced,
the current data is yet to show a clear evidence for non-vanishing isocurvature perturbations.
We suggest that precise measurements of the CMB polarization in the future are promising
to distinguish the anti-correlated isocurvature model from the adiabatic one.

\section*{Acknowledgments}

We would like to thank Jussi Valiviita and Matti Savelainen for valuable discussions and suggestions.
The work is supported by Grant-in-Aid for Scientific Research 25400248~(MK), 23740195~(TT), 24-2775~(SY)
and 21111006~(MK) from the Ministry of Education, Culture, Sports, Science and Technology in Japan,
and also by World Premier International Research Center Initiative (WPI Initiative), MEXT, Japan.
TS is supported by the Academy of Finland grant 1263714. 
We thank the CSC - IT Center for Science (Finland) for computational resources.

\end{document}